\documentclass[aps,twocolumn,nofootinbib]{revtex4}%
\usepackage{amsfonts}
\usepackage{amsmath}
\usepackage{amssymb}
\usepackage{graphicx}
\usepackage[usenames]{color}%
\providecommand{\U}[1]{\protect\rule{.1in}{.1in}}

\begin{document}
\preprint{ }
\title[Interface superconductivity]{Interface superconductivity in LaAlO$_{3}$-SrTiO$_{3}$ heterostructures}
\author{S. N. Klimin}
\thanks{On leave of absence from: Department of Theoretical Physics, State University
of Moldova, str. A. Mateevici 60, MD-2009 Kishinev, Republic of Moldova.}
\author{J. Tempere}
\thanks{Also at Lyman Laboratory of Physics, Harvard University, Cambridge, MA 02138, USA.}
\affiliation{Theorie van Kwantumsystemen en Complexe Systemen (TQC), Universiteit
Antwerpen, Universiteitsplein 1, B-2610 Antwerpen, Belgium}
\author{D. van der Marel}
\affiliation{D\'{e}partement de Physique de la Mati\`{e}re Condens\'{e}e, Universit\'{e} de
Gen\`{e}ve, CH-1211 Gen\`{e}ve 4, Switzerland}
\author{J. T. Devreese}
\affiliation{Theorie van Kwantumsystemen en Complexe Systemen (TQC), Universiteit
Antwerpen, Universiteitsplein 1, B-2610 Antwerpen, Belgium}

\begin{abstract}
The interface superconductivity in LaAlO$_{3}$-SrTiO$_{3}$ heterostructures
reveals a non-monotonic behavior of the critical temperature as a function of
the two-dimensional density of charge carriers. We develop a theoretical
description of interface superconductivity in strongly polar heterostructures,
based on the dielectric function formalism. The density dependence of the
critical temperature is calculated accounting for all phonon branches
including different types of optical (interface and half-space) and acoustic
phonons. The LO- and acoustic-phonon-mediated electron-electron interaction is
shown to be the dominating mechanism governing the superconducting phase
transition in the heterostructure.

\end{abstract}
\date{\today}
\maketitle

\section{Introduction}

Recent progress in the development of multilayer structures based on complex
oxides \cite{Ohtomo}, provides the means to generate a two-dimensional
electron gas (2DEG) at the oxide interfaces. The discovery of
superconductivity at the LaAlO$_{3}$-SrTiO$_{3}$ interface
\cite{Reyren,Reyren2,Caviglia,Schneider,Richter} has stimulated increasing
interest in the experimental and theoretical study of these structures.

Because strontium titanate is a highly polar crystal, the electron-phonon
mechanism of superconductivity seems to be the most promising for the
explanation of the experimental data on superconductivity in the LaAlO$_{3}%
$-SrTiO$_{3}$ heterostructures. The Migdal -- Eliashberg theory of
superconductivity \cite{Eliashberg,Scalapino}, as well the BCS theory, is
valid when the phonon frequencies are much smaller than the electron Fermi
energy. This is not the case for polar crystals with sufficiently high
optical-phonon frequencies, like strontium titanate. To tackle such systems,
non-adiabatic extensions of the theory of superconductivity have been
developed. Pietronero \emph{et al}. \cite{Pietronero,Pietronero1,Pietronero2}
generalized the Eliashberg equations to include non-adiabatic corrections
beyond Migdal's theorem. The method developed by Kirzhnits \emph{et al}.
\cite{Kirzhnits} (see also Refs. \cite{Takada1,Takada2,Takada3}) is focused on
the superconductivity caused by the Fr\"{o}hlich electron-phonon interaction
with polar optical phonons. It uses the total dielectric function of a polar
crystal. The Pietronero and Kirzhnits approaches are complementary: the former
is non-perturbative with respect to the coupling strength and perturbative
with respect to the Debye energy, while the latter is weak-coupling but
non-perturbative with respect to the optical-phonon energies.

Strontium titanate is a unique example of a polar medium in which
superconductivity has been detected at very low carrier densities, so that the
optical-phonon energies can be larger than the Fermi energy. Moreover, as
found in Refs. \cite{Mechelen2008,PRB2010}, the electron -- LO-phonon coupling
constant in SrTiO$_{3}$ is not very large. Therefore for the investigation of
superconductivity in a LaAlO$_{3}$-SrTiO$_{3}$ heterostructure, the Kirzhnits
method seems to be appropriate. Here, we apply the Kirzhnits method for a
multilayer structure with several polar layers.

\section{Superconductivity in a multilayer polar structure}

We consider the quasi 2D electron-phonon system described by the Hamiltonian:%
\begin{align}
H  &  =\sum_{\mathbf{k}_{\parallel}}\sum_{\sigma,j}\epsilon_{j,n,\mathbf{k}%
_{\parallel}}c_{\sigma,j,n,\mathbf{k}_{\parallel}}^{+}c_{\sigma,j,n,\mathbf{k}%
_{\parallel}}\nonumber\\
&  +\frac{1}{2L^{2}}\sum_{\mathbf{k}_{\parallel},\mathbf{k}_{\parallel
}^{\prime}}\sum_{\mathbf{q}}\sum_{\sigma,j,n,\sigma^{\prime},j^{\prime
},n^{\prime}}U_{\mathbf{q}}^{\left(  n^{\prime},n\right)  }\nonumber\\
&  \times c_{\sigma,j,n,\mathbf{k}_{\parallel}+\mathbf{q}}^{+}c_{\sigma
^{\prime},j^{\prime},n^{\prime},\mathbf{k}_{\parallel}^{\prime}}^{+}%
c_{\sigma^{\prime},j^{\prime},n^{\prime},\mathbf{k}_{\parallel}^{\prime
}\mathbf{+q}}c_{\sigma,j,n,\mathbf{k}_{\parallel}}\nonumber\\
&  +\sum_{\mathbf{q}}\sum_{\lambda}\hbar\Omega_{\mathbf{q},\lambda
}a_{\mathbf{q},\lambda}^{+}a_{\mathbf{q},\lambda}\nonumber\\
&  +\frac{1}{L}\sum_{\mathbf{q},\lambda}\left(  \gamma_{\mathbf{q},\lambda
}a_{\mathbf{q},\lambda}+\gamma_{\mathbf{q},\lambda}^{+}a_{\mathbf{q},\lambda
}^{+}\right)  . \label{H}%
\end{align}
Here $a_{\mathbf{q},\lambda},a_{\mathbf{q},\lambda}^{+}$ are the phonon second
quantization operators, $\mathbf{q}$ is the 2D in-plane phonon wave vector,
the index $\lambda$ labels phonon branches in the LAO-STO structure,
$\Omega_{\lambda}\left(  \mathbf{q}\right)  $ is the phonon frequency.
Furthermore, $c_{\sigma,j,n,\mathbf{k}_{\parallel}}^{+}$ and $c_{\sigma
,j,n,\mathbf{k}_{\parallel}}$ are, respectively, the creation and annihilation
operators for electrons with spin $\sigma$, in-plane wave vector
$\mathbf{k}_{\parallel}$, band index $j$ and size-quantization quantum number
$n$. The energy corresponding to the single-particle state $\left\vert
j,n,\mathbf{k}_{\parallel}\right\rangle $ is $\epsilon_{j,n,\mathbf{k}%
_{\parallel}}$. $L$ is the lateral size of the system, and $U_{\mathbf{q}%
}^{\left(  n^{\prime},n\right)  }$ is the matrix element of the
electron-electron interaction potential,%
\begin{align}
U_{\mathbf{q}}^{\left(  n^{\prime},n\right)  }  &  =\int dz\int dz^{\prime
}\tilde{U}_{C}\left(  q,z,z^{\prime}\right) \nonumber\\
&  \times\varphi_{n}\left(  z\right)  \varphi_{n^{\prime}}\left(  z\right)
\varphi_{n}\left(  z^{\prime}\right)  \varphi_{n^{\prime}}\left(  z^{\prime
}\right)  . \label{Uq}%
\end{align}
The electron-phonon interaction amplitudes can be written as:%
\begin{equation}
\gamma_{\mathbf{q},\lambda}=\sum_{j,n,j^{\prime},n^{\prime}}\Gamma
_{\mathbf{q},\lambda}^{\left(  n^{\prime},n\right)  }\sum_{\sigma}%
c_{\sigma,j^{\prime},n^{\prime},\mathbf{k}_{\parallel}+\mathbf{q}}%
^{+}c_{\sigma,j,n,\mathbf{k}_{\parallel}}. \label{gamma}%
\end{equation}
where $\Gamma_{\mathbf{q},\lambda}^{\left(  n^{\prime},n\right)
}=\left\langle \varphi_{n^{\prime}}\left\vert \Gamma_{\lambda}\left(
\mathbf{q},z\right)  \right\vert \varphi_{n}\right\rangle $ is the matrix
element of the amplitude $\Gamma_{\lambda}\left(  \mathbf{q},z\right)  $ that
we will specify below. The index $\lambda$ labels the phonon branches of the
multilayer structure. In the calculations, we assume that in the LaAlO$_{3}%
$-SrTiO$_{3}$ heterostructure under consideration, the electron gas is
confined to a very thin layer $\sim$ 2 nm. Consequently, only the lowest
energy subband ($n=0$) is filled, and transitions to higher subbands can be neglected.

The electron-electron interaction potential $\tilde{U}_{C}\left(
q,z,z^{\prime}\right)  $, the equations for the eigenfrequencies of the
interface modes, and the amplitudes of the electron-phonon interaction in a
multilayer structure are derived within the dielectric continuum approach
accounting for the electrostatic boundary conditions in a similar way as in
Refs. \cite{Mori1989,Hai1990}. We use Feynman units: $\hbar=1$, $m_{b}=1$,
$\omega_{0}=1$, where $\omega_{0}$ is an effective LO-phonon frequency (taken
equal to the highest of the LO-phonon frequencies of SrTiO$_{3}$). The
potential $\tilde{U}_{C}\left(  q,z,z^{\prime}\right)  $ is%
\begin{equation}
\tilde{U}_{C}\left(  q,z,z^{\prime}\right)  =\frac{1}{L}\frac{2\sqrt{2}%
\pi\alpha_{0}}{\varepsilon_{1,\infty}q}\left[  e^{-q\left\vert z-z^{\prime
}\right\vert }+C_{I}e^{q\left(  z+z^{\prime}\right)  }\right]  , \label{Uc2}%
\end{equation}
expressed using the dimensionless Coulomb coupling constant%
\begin{equation}
\alpha_{0}=\frac{e^{2}}{2\hbar\omega_{0}}\left(  \frac{2m_{b}\omega_{0}}%
{\hbar}\right)  ^{1/2}. \label{alpha0}%
\end{equation}
The coefficient $C_{I}$ depends on the dielectric constants of the media
constituting the heterostructure. For the system without an electrode at the
LaAlO$_{3}$-vacuum interface, $C_{I}$ is%
\begin{equation}
C_{I}=\frac{\varepsilon_{1,\infty}\varepsilon_{3,\infty}-\varepsilon
_{2,\infty}^{2}+\varepsilon_{2,\infty}\left(  \varepsilon_{1,\infty
}-\varepsilon_{3,\infty}\right)  \coth\left(  ql\right)  }{\varepsilon
_{1,\infty}\varepsilon_{3,\infty}+\varepsilon_{2,\infty}^{2}+\varepsilon
_{2,\infty}\left(  \varepsilon_{1,\infty}+\varepsilon_{3,\infty}\right)
\coth\left(  ql\right)  }, \label{CI1}%
\end{equation}
where $l$ is the width of the LaAlO$_{3}$ layer. The index $s=1,2,3$ in the
dielectric constant $\varepsilon_{s,\infty}$ labels the layers: $s=1$ for the
SrTiO$_{3}$ substrate, $s=2$ for the LaAlO$_{3}$ layer, and $s=3$ for the
vacuum. For the system with an electrode, $C_{I}$ is obtained from (\ref{CI1})
in the limit $\varepsilon_{3,\infty}\rightarrow\infty$.

We take into account the following phonon branches: (1) the interface optical
phonons, (2) the half-space optical phonons, and (3) the acoustic phonons. For
the interface optical phonons, the eigenfrequencies are found from the
equation
\begin{equation}
\nu_{1}\left(  \Omega_{\lambda}\right)  \nu_{2}\left(  \Omega_{\lambda
}\right)  -\mu_{2}^{2}\left(  \Omega_{\lambda}\right)  =0 \label{eqs}%
\end{equation}
with the functions%
\begin{align}
\nu_{1}\left(  \Omega_{\lambda}\right)   &  =\varepsilon_{1}\left(
\Omega_{\lambda}\right)  +\varepsilon_{2}\left(  \Omega_{\lambda}\right)
\coth\left(  ql\right)  ,\label{n1}\\
\nu_{2}\left(  \Omega_{\lambda}\right)   &  =\varepsilon_{2}\left(
\Omega_{\lambda}\right)  \coth\left(  ql\right)  +\varepsilon_{3}\left(
\Omega_{\lambda}\right)  ,\label{n2}\\
\mu_{2}\left(  \Omega_{\lambda}\right)   &  =\frac{\varepsilon_{2}\left(
\Omega_{\lambda}\right)  }{\sinh\left(  ql\right)  }. \label{m}%
\end{align}
The amplitudes of the electron-phonon interaction with these interface phonon
modes are%
\begin{align}
\Gamma_{\lambda}\left(  \mathbf{q},z\right)   &  =\left(  2\sqrt{2}\pi
\alpha_{0}\right)  ^{1/2}\left(  \frac{1}{q}\frac{\Omega_{\lambda}}{D\left(
\Omega_{\lambda}\right)  }\right)  ^{1/2}\nonumber\\
&  \times\left[  e^{qz}\Theta\left(  -z\right)  +\frac{\nu_{1}\left(
\Omega_{\lambda}\right)  }{\mu_{2}\left(  \Omega_{\lambda}\right)
}e^{q\left(  l-z\right)  }\Theta\left(  z-l\right)  \right. \nonumber\\
&  +\Theta\left(  z\right)  \Theta\left(  l-z\right) \nonumber\\
&  \left.  \times\left(  \frac{\sinh\left[  q\left(  l-z\right)  \right]
}{\sinh\left(  ql\right)  }+\frac{\nu_{1}\left(  \Omega_{\lambda}\right)
}{\mu_{2}\left(  \Omega_{\lambda}\right)  }\frac{\sinh\left(  qz\right)
}{\sinh\left(  ql\right)  }\right)  \right]  \label{G3}%
\end{align}
where $\Theta\left(  z\right)  $ is the Heaviside step function, and the
factor $D\left(  \Omega_{\lambda}\right)  $ is%
\begin{align}
D\left(  \Omega_{\lambda}\right)   &  =\left(  \varepsilon_{1,0}%
-\varepsilon_{1,\infty}\right)  \left(  \frac{\Omega_{\lambda}\omega_{1,TO}%
}{\Omega_{\lambda}^{2}-\omega_{1,TO}^{2}}\right)  ^{2}\nonumber\\
&  +\left(  \varepsilon_{2,0}-\varepsilon_{2,\infty}\right)  \left(
\frac{\Omega_{\lambda}\omega_{2,TO}}{\Omega_{\lambda}^{2}-\omega_{2,TO}^{2}%
}\right)  ^{2}\nonumber\\
&  \times\frac{\left(  \left(  \frac{\nu_{1}\left(  \Omega_{\lambda}\right)
}{\mu_{2}\left(  \Omega_{\lambda}\right)  }\right)  ^{2}+1\right)
\cosh\left(  ql\right)  -2\frac{\nu_{1}\left(  \Omega_{\lambda}\right)  }%
{\mu_{2}\left(  \Omega_{\lambda}\right)  }}{\sinh\left(  ql\right)
}\nonumber\\
&  +\left(  \varepsilon_{3,0}-\varepsilon_{3,\infty}\right)  \left(
\frac{\Omega_{\lambda}\omega_{3,TO}}{\Omega_{\lambda}^{2}-\omega_{3,TO}^{2}%
}\right)  ^{2}\left(  \frac{\nu_{1}\left(  \Omega_{\lambda}\right)  }{\mu
_{2}\left(  \Omega_{\lambda}\right)  }\right)  ^{2}. \label{D1}%
\end{align}
For the structure with an electrode, we set $\varepsilon_{3}\left(
\omega\right)  \rightarrow\infty$ in the above formulae.

Because the 2DEG layer is positioned at the SrTiO$_{3}$ side of the interface,
the half-space phonons of strontium titanate can contribute to
superconductivity. The frequencies of the half-space phonons are the same as
for the bulk LO phonons. The amplitudes of the electron-phonon interaction for
the half-space phonons differ from those for the bulk LO phonons only by the
boundary condition of zero amplitude at the interface. Although the half-space
phonons turn out to give a relatively very small contribution to the resulting
phonon-mediated interaction potential, we take them into account for
completeness. For the acoustic-phonon contribution, we use the frequencies and
interaction amplitudes for the deformation potential from Ref. \cite{PD1985}:%
\begin{align}
\omega_{\mathbf{q}}  &  =vq,\label{Vqac}\\
V_{\mathbf{q}}^{\left(  ac\right)  }  &  =\left(  4\pi\alpha_{ac}\right)
^{1/2}\frac{\hbar^{2}}{m_{b}}q^{1/2}%
\end{align}
with the dimensionless coupling constant%
\begin{equation}
\alpha_{ac}=\frac{D^{2}m_{b}^{2}}{8\pi\rho\hbar^{3}v}, \label{alphac}%
\end{equation}
where $\rho$ is the mass density of strontium titanate, $D$ is the deformation
potential, and $v$ is the sound velocity.

The calculation of the superconducting transition temperature is performed
following the scheme of Refs. \cite{Kirzhnits,Takada1,Takada2,Takada3} using
the gap equation%
\begin{equation}
\Delta\left(  \omega\right)  =-\int_{-\epsilon_{F}}^{\infty}\frac
{d\omega^{\prime}}{2\omega^{\prime}}\tanh\left(  \frac{\beta\omega^{\prime}%
}{2}\right)  \Delta\left(  \omega^{\prime}\right)  K\left(  \omega
,\omega^{\prime}\right)  \label{Gap4}%
\end{equation}
with the kernel function%
\begin{align}
K\left(  \omega,\omega^{\prime}\right)   &  =\frac{m_{b}}{\pi^{3}}\int
_{0}^{\pi}d\varphi\int_{0}^{\infty}d\Omega\frac{\left\vert \omega\right\vert
+\left\vert \omega^{\prime}\right\vert }{\Omega^{2}+\left(  \left\vert
\omega\right\vert +\left\vert \omega^{\prime}\right\vert \right)  ^{2}%
}\nonumber\\
&  \times V^{tot}\left(  q,i\Omega\right)  , \label{K1}%
\end{align}
where $q=\sqrt{p^{2}+k^{2}-2pk\cos\varphi},$ $p=\sqrt{2m_{b}\left(
\omega+\epsilon_{F}\right)  },$ $k=\sqrt{2m_{b}\left(  \omega^{\prime
}+\epsilon_{F}\right)  }$, $m_{b}$ is the effective mass for the motion along
the surface, and $V^{tot}\left(  q,i\Omega\right)  $ is the total effective
electron-electron interaction potential. The energy $\omega$ is counted from
the Fermi energy $\epsilon_{F}$. The kernel function (\ref{K1}) is essentially
energy-nonlocal, as distinct from the BCS and Migdal -- Eliashberg approaches,
since it is provided by a retarded effective electron-electron interaction
$V^{tot}\left(  q,i\Omega\right)  $, through the plasmon-phonon excitations.
Consequently, the frequency dependence of the gap $\Delta\left(
\omega\right)  $ can differ from that within the BCS or Migdal-Eiashberg pictures.

The gap equation (\ref{Gap4}) with the effective interaction potential
described above allows for the determination of the gap function
$\Delta\left(  \omega\right)  $ and the critical temperature in a LaAlO$_{3}%
$-SrTiO$_{3}$ heterostructure. In the low-temperature range, when the thermal
energy $k_{B}T$ is much lower than the Fermi energy of the charge carriers
$\epsilon_{F}$, the approximation method proposed by Zubarev \cite{Zubarev}
allows to find the normalized gap function $\phi\left(  \omega\right)
\equiv\Delta\left(  \omega\right)  /\Delta\left(  0\right)  $ as a numeric
solution of the Fredholm equation,%
\begin{align}
&  \phi\left(  \omega\right)  +\int_{-\epsilon_{F}}^{\infty}d\omega^{\prime
}~\phi\left(  \omega^{\prime}\right)  \nonumber\\
&  \times\frac{1}{2\left\vert \omega^{\prime}\right\vert }\left[  K\left(
\omega,\omega^{\prime}\right)  -\frac{K\left(  \omega,0\right)  K\left(
0,\omega^{\prime}\right)  }{K\left(  0,0\right)  }\right]  \nonumber\\
&  =\frac{K\left(  \omega,0\right)  }{K\left(  0,0\right)  }.\label{phi}%
\end{align}
The critical temperature is given by the expression,%
\begin{equation}
T_{c}=\frac{2}{\pi}e^{\gamma}\epsilon_{F}\exp\left(  -\frac{1}{\lambda
}\right)  \approx1.14\epsilon_{F}\exp\left(  -\frac{1}{\lambda}\right)
,\label{Tc}%
\end{equation}
where $\gamma=0.577216\ldots$ is the Euler constant, and the parameter
$\lambda$ is determined explicitly through the normalized gap parameter%
\begin{align}
\frac{1}{\lambda}  & =-\left\{  \frac{1}{K\left(  0,0\right)  }+\int
_{-\epsilon_{F}}^{\infty}\frac{d\omega}{2\left\vert \omega\right\vert
}\right.  \nonumber\\
& \left.  \times\left[  \frac{K\left(  0,\omega\right)  }{K\left(  0,0\right)
}\phi\left(  \omega\right)  -\Theta\left(  \epsilon_{F}-\omega\right)
\right]  \right\}  .\label{lamb}%
\end{align}
with the Heaviside step function $\Theta\left(  \epsilon_{F}-\omega\right)  $.
Formulae (\ref{phi}) and (\ref{Tc}) describe the relation between the kernel
function $K\left(  \omega,\omega^{\prime}\right)  $, the normalized gap
function $\phi\left(  \omega\right)  $ and the critical temperature.

In the present treatment, the effective electron-electron interaction includes
contributions from both optical and acoustic phonons. Since the Kirzhnits
theory assumes the weak-coupling regime, we suggest that the effective
phonon-mediated interaction due to the acoustic phonons can be taken into
account in an additive way with respect to the combined contribution of
Coulomb interaction and optical phonons. The total effective interaction
$V^{tot}\left(  q,\Omega\right)  $ can be thus approximated by the sum:%
\begin{equation}
V^{tot}\left(  q,i\Omega\right)  =V^{R}\left(  q,i\Omega\right)
+V^{ac}\left(  q,i\Omega\right)  , \label{Vtot1}%
\end{equation}
where $V^{R}\left(  q,\Omega\right)  $ is the effective interaction described
in terms of the total dielectric function, and $V^{ac}\left(  q,i\Omega
\right)  $ is the effective interaction due to the acoustic phonons.

The effective potential $V^{R}\left(  q,i\Omega\right)  $ in a quasi 2D system
is determined following Ref. \cite{Takada3}. Within RPA \cite{PN}, the
relation between the effective potential taking into account dynamic
screening, $V^{R}\left(  q,i\Omega\right)  $, and the effective potential
without screening, $V_{0}^{R}\left(  q,i\Omega\right)  $, is%
\begin{equation}
V^{R}\left(  q,i\Omega\right)  =\frac{V_{0}^{R}\left(  q,i\Omega\right)
}{1+V_{0}^{R}\left(  q,i\Omega\right)  P^{\left(  1\right)  }\left(
q,i\Omega\right)  } \label{RPA}%
\end{equation}
where $P^{\left(  1\right)  }\left(  q,i\Omega\right)  $ is the polarization
function of a free 2DEG. Here, we use the RPA-polarization function
\cite{PN,Takada1,Ando}. The non-screened potential $V_{0}^{R}\left(
q,i\Omega\right)  $ is a sum of a Coulomb contribution and a contribution from
the phonon mediated interaction between the two electrons:%
\begin{equation}
V_{0}^{R}\left(  q,i\Omega\right)  =U_{\mathbf{q}}^{\left(  0,0\right)  }%
-\sum_{\lambda}\frac{2\Omega_{\lambda}\left(  q\right)  }{\Omega_{\lambda}%
^{2}\left(  q\right)  +\Omega^{2}}\left\vert \Gamma_{\mathbf{q},\lambda
}^{\left(  0,0\right)  }\right\vert ^{2} \label{VEff}%
\end{equation}
where the Coulomb contribution $U_{\mathbf{q}}^{\left(  0,0\right)  }$ is
given by the expression (\ref{Uq}), and the effective optical phonon mediated
interaction is approximated by the Bardeen -- Pines form \cite{BP,Kelly}. For
the acoustic-phonon contribution to the effective potential, we also apply the
Bardeen-Pines approximation, as in Ref. \cite{PRB2012}:%
\begin{equation}
V^{ac}\left(  q,i\Omega\right)  =-\frac{1}{\hbar}\left\vert V_{\mathbf{q}%
}^{\left(  ac\right)  }\right\vert ^{2}\frac{2\omega_{\mathbf{q}}}%
{\omega_{\mathbf{q}}^{2}+\Omega^{2}}. \label{Vac}%
\end{equation}

There are indications from experiment \cite{Reyren,Reyren2,Caviglia,Schneider}
that the superconducting phase transition in a LaAlO$_{3}$-SrTiO$_{3}$
heterostructure is governed by the Berezinskii -- Kosterlitz -- Thouless (BKT)
mechanism \cite{Berezinskii,KT,K2}. It is shown in Refs.
\cite{Caviglia,Schneider} that the temperature dependence of the resistance
just above $T_{c}$ is specific for a BKT phase transition, corresponding to
the 2D nature of the superconducting system.

The critical temperature of the BKT phase transition is determined by the
equation \cite{KT} which includes the pair superfluid density $\rho_{s}\left(
T\right)  $:%
\begin{equation}
T_{BKT}=\frac{\pi}{4}\frac{\hbar^{2}}{k_{B}m_{b}}\rho_{s}\left(
T_{BKT}\right)  . \label{TBKT}%
\end{equation}
The superfluid density monotonously decreases with increasing temperature, and
turns to zero at $T=T_{c}$. Therefore, the critical temperature $T_{BKT}$ must
be necessarily lower than $T_{c}$. In the case when the BKT transition is
present in the LaAlO$_{3}$-SrTiO$_{3}$ heterostructure, $T_{c}$ can be
interpreted as the pairing temperature at which the preformed pairs appear. In
the LaAlO$_{3}$-SrTiO$_{3}$ heterostructures, the superfluid density $n_{s}$
extracted from the BKT equation (\ref{TBKT}) is several orders of magnitude
lower than the actual electron density: $n_{s}\ll n_{0}$. This inequality can
be satisfied only when the gap parameter $\Delta$ is very small compared to
its value at $T=0$, and, consequently, when $\left(  1-T_{BKT}/T_{c}\right)
\ll1$. Therefore, as already concluded in Ref. \cite{Reyren}, $T_{c}$ and
$T_{BKT}$ are extremely close to each other in the LaAlO$_{3}$-SrTiO$_{3}$ heterostructures.

\section{Results and discussion}

For the numerical calculations, we use the set of material parameters already
used in earlier works \cite{PRB2010,PRB2012}. The dielectric constants for
SrTiO$_{3}$ are $\varepsilon_{1,\infty}=5.44$ and $\varepsilon_{1,0}=186$
(calculated using the Lyddane-Sachs-Teller relation for the LO- and TO- phonon
frequencies and the ratio $\varepsilon_{0}/\varepsilon_{\infty}$). The
effective mass for the present calculation has been taken $m_{b}=1.65m_{0}$
\cite{PRB2010} (where $m_{0}$ is the electron mass in vacuum). The dielectric
constants for LaAlO$_{3}$ are used from Ref. \cite{Calvani}: $\varepsilon
_{2,\infty}=4.2$ and $\varepsilon_{2,0}=24$. The only material parameter which
is not yet well-determined, is the acoustic deformation potential $D$ in
strontium titanate. It should be noted that the deformation potential
responsible for the interaction of an electron with the acoustic phonons is
the \textquotedblleft absolute\textquotedblright\ rather than
\textquotedblleft relative\textquotedblright\ deformation potential
\cite{Cardona,VDW1,Zunger1994}. In the literature, we can find several
different suggestions on the values of the deformation potential in strontium
titanate. Koonce \emph{et al}. \cite{Koonce} applied the value $D\approx15$ eV
to fit the experimental data on $T_{c}$ in bulk strontium titanate. In Ref.
\cite{Morozovska} the deformation potential is estimated to be $D\approx2.9$
eV on the basis of the value of the Fermi energy of the electrons. In Ref.
\cite{Janotti2} the value $D\approx4$ eV is calculated on the basis of first
principles density functional theory, that seems to be more reliable than two
other values, because the many-valley band model of Ref. \cite{Koonce} is not
confirmed by later studies, and the deformation potential of Ref.
\cite{Morozovska} is a rough estimation using the Fermi energy of the
electrons. As we discuss below, the results from our theory compare favorably
with $D$ values of \cite{Morozovska,Janotti2}, but are incompatible with the
large $D$ value used in \cite{Koonce}.

Here we have calculated $T_{c}$ in the LaAlO$_{3}$-SrTiO$_{3}$ heterostructure
using several values of the deformation potential: $D=3%
\operatorname{eV}%
$, $D=4%
\operatorname{eV}%
$ and $D=5%
\operatorname{eV}%
$. They seem to be physically reasonable, because they lie in the same range
as the values used in Refs. \cite{Morozovska,Janotti2}. For the comparison of
the calculated critical temperatures with the known experimental data, we use
in the numeric calculations the model of the LaAlO$_{3}$-SrTiO$_{3}$
heterostructure accounting for the presence of an electrode at the oxide layer.

In Fig. 1, the kernel function $K\left(  \omega,\omega^{\prime}\right)  $ is
plotted for the set of parameters indicated above, choosing the deformation
potential $D=4$ eV suggested in Ref. \cite{Janotti2}. This kernel function is
qualitatively similar to the kernel function for a 2D electron gas from Ref.
\cite{Takada2}. There exists a distinction between the kernel functions for
the 2D and 3D systems within the Kirzhnits -- Takada method: for a 3D electron
gas, the kernel function $K\left(  0,\omega\right)  $ tends to zero when
$\omega\rightarrow-\epsilon_{F}$ achieving a local maximum in the interval
$-\epsilon_{F}<\omega<0$. On the contrary, for a 2D electron gas, $K\left(
\omega,\omega^{\prime}\right)  $ is a monotonically decreasing function of
$\omega$ and $\omega^{\prime}$ in the range of negative frequencies.%

\begin{figure}
[h]
\begin{center}
\includegraphics[
height=2.2208in,
width=2.9456in
]%
{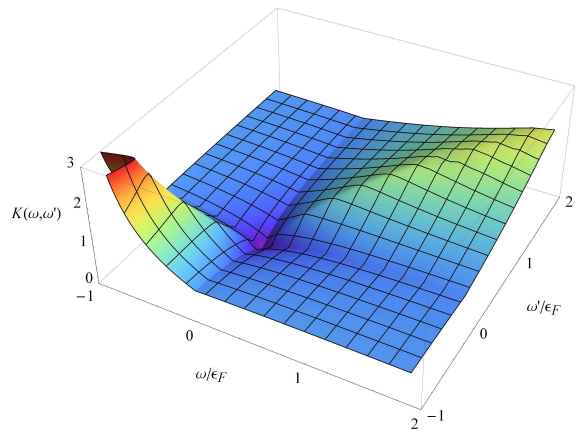}%
\caption{Kernel function $K\left(  \omega,\omega^{\prime}\right)  $ for the
LaAlO$_{3}$-SrTiO$_{3}$ structure using the set of parameters described in the
text.}%
\end{center}
\end{figure}

As explained in Refs. \cite{Takada1,Takada2,Takada3}, superconductivity in an
electron-phonon system can exist despite the fact that the kernel function
$K\left(  \omega,\omega^{\prime}\right)  $ is positive for all frequencies.
The kernel for energies larger than $\epsilon_{F}$ is dominated by the Coulomb
repulsion between two electrons whose spatial distance is small, while the
behavior of the kernel near the Fermi surface is due to both the Coulomb
interaction and the attraction mediated by the plasmon-phonon excitations,
between two electrons whose distance is rather large. Consequently, when, for
example, $K\left(  \epsilon_{F},0\right)  $ is much larger than $K\left(
0,0\right)  $, the two electrons can avoid the region of the Coulomb repulsion
and form a Cooper pair. We can see that, although $K\left(  0,0\right)  $ is
not exactly equal to zero, $K\left(  \omega,\omega^{\prime}\right)  $ achieves
its minimum at $\omega,\omega^{\prime}=0$, facilitating pairing.

The measured critical temperatures are taken from several sources
\cite{Reyren,Caviglia,Richter}. The experimental work by N. Reyren \emph{et
al}. \cite{Reyren} contains only two points: $T_{c}\approx0.1$~K for a 2D
density of the electrons $n\approx1.5\times10^{13}~$cm$^{-2}$, and
$T_{c}\approx0.2$~K for $n\approx4\times10^{13}~$cm$^{-2}$. The paper
\cite{Caviglia} represents the critical temperature as a function of the gate
voltage, and the dependence of the modulation of the electron density $\delta
n\left(  V\right)  $ on the gate voltage. The total electron density is
related to the modulation $\delta n\left(  V\right)  $ as $n\left(  V\right)
=n_{0}+\delta n\left(  V\right)  $, where $n_{0}\approx4.5\times10^{13}%
~$cm$^{-2}$, according to Ref. \cite{Caviglia}. Using the experimental data
for $T_{c}\left(  V\right)  $ and $n\left(  V\right)  $ represented in these
figures, we obtain the dependence $T_{c}\left(  n\right)  $ for the experiment
\cite{Caviglia}. We also include recent experimental results on the
superconductivity in the LaAlO$_{3}$-SrTiO$_{3}$ heterostructure
\cite{Richter}.

In Fig. 2 (\emph{a}), the critical temperatures as a function of the 2D
carrier density calculated in the present work are compared to the
experimental data for $T_{c}\left(  n\right)  $ discussed above. It is worth
noting that there exists a substantial difference between experimental values
of $T_{c}$ obtained in different experiments. However, they all are of the
same order of magnitude and lie in the same range of the carrier densities.%

\begin{figure}
[h]
\begin{center}
\includegraphics[
height=4.2445in,
width=2.9118in
]%
{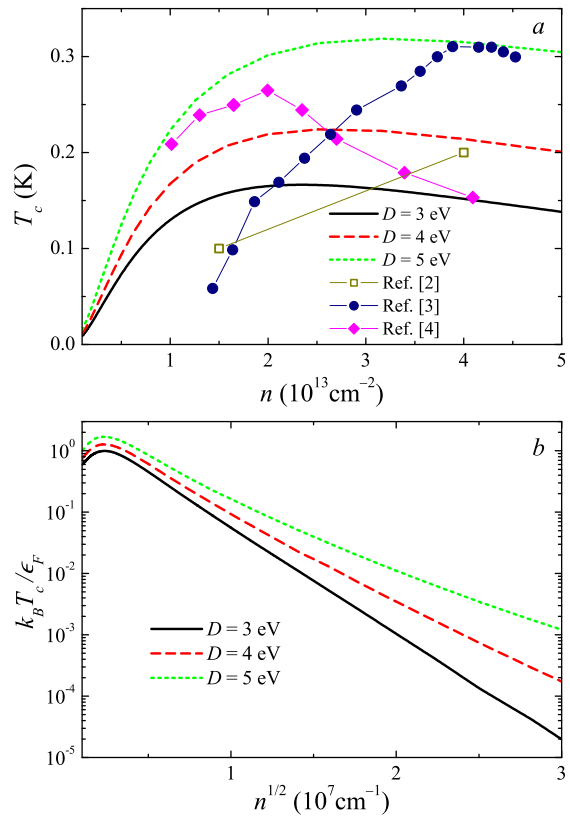}%
\caption{(\emph{a}) Critical temperature for the LaAlO$_{3}$-SrTiO$_{3}$
heterostructure as a function of the 2D electron density, compared to the
experimental data extracted from Refs. \cite{Reyren,Caviglia,Richter}.
(\emph{b}) The calculated critical temperatures divided by the Fermi energy
$\epsilon_{F}$, plotted as a function of $n^{1/2}$.}%
\end{center}
\end{figure}

The observed differences of the experimental results on the critical
temperatures in the LaAlO$_{3}$-SrTiO$_{3}$ heterostructures can be explained
as follows. In different experiments \cite{Reyren,Caviglia,Richter}, the
LaAlO$_{3}$-SrTiO$_{3}$ heterostructures have been grown independently.
Therefore those heterostructures can at least slightly differ from each other.
The thermal energy $k_{B}T_{c}$ is extremely small compared to the
characteristic energies involved in the superconducting phase transition: the
Fermi energy of the electrons and the LO-phonon energies (which both are of
order $\sim100$ meV). Under these conditions, the critical temperatures can be
very sensitive to relatively small difference of the internal properties of
the fabricated heterostructures. Additional factors (e. g., disorder, local
phonons, defects, etc.) can substantially influence $T_{c}$. Moreover, also
even the critical temperatures for bulk strontium titanate measured in
different experiments \cite{Koonce,Binnig} differ substantially from each
other. Consequently, the relatively large variation of the experimental
results on the critical temperature in LaAlO$_{3}$-SrTiO$_{3}$ heterostructure
is not surprising. Despite that uncertainty, the measured critical
temperatures in different experiments are of the same order.

We can see that for each $D$, the curve is close to only few data points.
However (i) experimental data are obtained with a substantial numeric
inaccuracy, (ii) experimental data from different sources do not agree even
with each other. It should be noted also that, since the thermal energy
corresponding to the critical temperature in the LAO-STO structure is very
small with respect to other energies participating in the superconductivity
(the optical-phonon energies and the Fermi energy), even a small uncertainty
of these parameters can then lead to a significant change of the critical
temperature. Our calculation, performed without fit using parameter values
known from literature, yields the critical temperatures within the same range
as in the experiments. We can therefore conclude that, taking into account the
uncertainty of the experimental results on the critical temperatures, the
suggested theoretical explanation of the superconducting phase transition in
the LaAlO$_{3}$-SrTiO$_{3}$ heterostructures leads to tentative agreement with experiment.

We can see different regimes of $T_{c}$ as a function of density in Fig. 2.
First, the critical temperature exhibits almost linear dependence at small $n$
up to $n\sim10^{13}%
\operatorname{cm}%
^{-2}$. After reaching the maximum, as seen from Fig. 2~(b), the critical
temperature falls down at high densities as $\sim n\exp\left(  -c\sqrt
{n}\right)  $ with a positive constant $c$ depending on the parameters of the
system. This density dependence of $T_{c}$ for low and high densities can be
explained as follows.

In the limit of low densities, the effective interaction potential
$V^{R}\left(  q,i\Omega\right)  $ given by (\ref{RPA}) tends to the
non-screened effective interaction potential $V_{0}^{R}\left(  q,i\Omega
\right)  $ that does not depend on the density. Also the contribution
$V^{ac}\left(  q,i\Omega\right)  $ to the total effective interaction due to
the acoustic phonons is density-independent. Therefore at low densities the
parameters $\lambda$ tends to a constant depending on the material parameters
of the system. According to (\ref{Tc}), the critical temperature in the
low-density limit is proportional to the carrier density.

This result has a clear physical interpretation. Let us compare (\ref{Tc})
with the known BCS expression \cite{BCS},%
\begin{equation}
T_{c}\approx1.14\hbar\omega_{D}\exp\left(  -\frac{1}{N\left(  \epsilon
_{F}\right)  V}\right)  , \label{BCS}%
\end{equation}
where $\omega_{D}$ is the Debye frequency, $N\left(  \epsilon_{F}\right)  $ is
the density of states at the Fermi energy, and $V$ is the model BCS matrix
element. The BCS theory describes the adiabatic regime when $\hbar\omega
_{D}\ll\epsilon_{F}$, and pairing occurs for the electrons whose energies lie
in the layer of width $\delta\varepsilon\sim\hbar\omega_{D}$ near the Fermi
energy. As a result, $T_{c}\propto\hbar\omega_{D}$ within the BCS picture. On
the contrary, in strontium titanate and in the LaAlO$_{3}$-SrTiO$_{3}$
heterostructure at low densities the anti-adiabatic regime is realized:
$\epsilon_{F}\ll\hbar\omega_{L,j}$, where $\omega_{L,j}$ is an optical-phonon
frequency. In the anti-adiabatic regime, all electrons participate in the
superconductivity. Therefore the factor $\hbar\omega_{D}$ in the adiabatic
regime corresponds to the factor $\epsilon_{F}$ in the anti-adiabatic regime.

Because in the anti-adiabatic regime all electrons contribute to
superconductivity, the parameter $\lambda$ hardly can be interpreted as
$N\left(  \epsilon_{F}\right)  V$. In a non-adiabatic regime $\lambda$ must
be, in general, a functional of the density of states for all energies
$0<\epsilon<\epsilon_{F}$. However, the density of states for a 2D system and
for a sufficiently low energy (where the band nonparabolicity is relatively
small) is%
\begin{equation}
N\left(  \epsilon\right)  =\frac{m_{b}}{\pi\hbar^{2}}, \label{N}%
\end{equation}
so that $N\left(  \epsilon\right)  $ (and hence also $\lambda$) does not
depend on the carrier concentration at low concentrations. Thus the aforesaid
qualitative physical estimation leads to the the low-density behavior
$T_{c}\left(  n\right)  \propto n$, in agreement with the result obtained in
the present work.

In the opposite regime of high carrier densities, the plasma frequency can
exceed both the Fermi energy and the optical phonon energies. In this regime,
the plasmon mechanism of superconductivity \cite{Takada2} must dominate.
According to Ref. \cite{Takada2}, the critical temperature for an electron gas
in 2D with the effective mass $m_{b}$ and the dielectric constant
$\varepsilon$ due to the plasmon mechanism can be modeled by an analytic
expression,%
\begin{equation}
T_{c}=\frac{2}{\pi}e^{\gamma}\epsilon_{F}\exp\left[  -\frac{\left(
1+\left\langle F\right\rangle \right)  ^{2}}{\left\langle F^{2}\right\rangle
-K\left(  0,0\right)  }\right]  \label{b1}%
\end{equation}
with the averages%
\begin{equation}
\left\langle A\right\rangle \equiv\int_{-\epsilon_{F}}^{\epsilon_{F}}%
\frac{d\omega}{2\left\vert \omega\right\vert }A\left(  \omega\right)
\label{b2}%
\end{equation}
and the function%
\begin{equation}
F\left(  \omega\right)  =\frac{1}{4\pi g_{v}}\sqrt{\frac{q_{TF}}{p_{F}}%
}B\left(  \frac{1}{4},\frac{1}{2}\right)  \sqrt{\frac{\left\vert
\omega\right\vert }{\epsilon_{F}}}. \label{b3}%
\end{equation}
Here, $B\left(  x,y\right)  $ is the Euler beta function, $g_{v}$ is the
conduction band degeneracy, $p_{F}$ is the Fermi momentum, and $q_{TF}$ is the
Thomas-Fermi wave vector. For a 2D electron gas, $q_{TF}=2g_{v}e^{2}%
m_{b}/\varepsilon$ does not depend on the carrier density. Here, the factor
$g_{v}$ is equal to 1 because the conduction band in SrTiO$_{3}$ is split due
to the spin-orbit interaction \cite{VDM2011}.

After the integration in (\ref{b2}), the critical temperature (\ref{b1}) takes
the form%
\begin{equation}
T_{c}=\frac{2}{\pi}e^{\gamma}\epsilon_{F}\exp\left[  -\frac{\left(
1+2C\right)  ^{2}}{C^{2}-K\left(  0,0\right)  }\right]  , \label{Tc2}%
\end{equation}
where $C$ is given by:%
\begin{equation}
C=\frac{1}{2\pi}B\left(  \frac{1}{4},\frac{1}{2}\right)  \sqrt{\frac{q_{TF}%
}{p_{F}}}. \label{C}%
\end{equation}
The upper bound for the density when $T_{c}=0$ is determined by the equation
\begin{equation}
C^{2}-K\left(  0,0\right)  =0. \label{crd}%
\end{equation}
The parameter $C$ is proportional to $n^{-1/4}$, and $C\ll1$ for sufficiently
high densities. Therefore in the high-density range, but for densities smaller
than that determined by (\ref{crd}), the model critical temperature (\ref{b1})
due to the plasmon mechanism behaves approximately as%
\begin{equation}
T_{c}\approx\frac{2}{\pi}e^{\gamma}\epsilon_{F}\exp\left(  -1.43554\frac
{p_{F}}{q_{TF}}\right)  . \label{aa}%
\end{equation}
Since $p_{F}=\sqrt{2\pi n}$, the estimation (\ref{aa}) following Ref.
\cite{Takada2} is in agreement with the critical temperature obtained in the
present work, as seen from Fig. 2~(b). In the figure, the ratio $k_{B}%
T_{c}/\epsilon_{F}$ is plotted in the logarithmic scale as a function of
$n^{1/2}$, focusing at the high-density range (larger than in the experiments
\cite{Reyren,Caviglia,Richter}). We can see that at relatively small acoustic
deformation potential $D=3%
\operatorname{eV}%
$, the dependence $\ln T_{c}$ as a function of $n^{1/2}$ is almost linear for
high densities. For larger $D$, the acoustic-phonon mechanism stronger
influences the density dependence of $T_{c}$ leading to deviations from the
purely plasmon picture.

\section{Conclusions}

In conclusion, we have re-formulated the Kirzhnits method for a multilayer
structure with several polar layers. The developed technique is capable to
describe superconductivity in multilayer structures, where the electrostatic
electron-electron interaction, the optical-phonon spectra, and the amplitudes
of the electron-phonon interaction are modified compared to bulk. In the
present treatment, all phonon branches existing in the multilayer structure
are taken into account.

We have found that at low densities, the critical temperature is well
described by a BCS-like expression with the Fermi energy instead of the Debye
energy. This is a direct consequence of the anti-adiabatic regime, which
occurs at low carrier densities. At high densities, the density dependence of
the critical temperature shows the domination of the plasmon mechanism of superconductivity.

The obtained agreement of the calculated critical temperatures with experiment
gives support to the hypothesis that the mechanism of superconductivity is
provided by the electron -- optical-phonon interaction (see, e. g.,
\cite{Dolgov}), at least in the multilayer structure analyzed in the present work.

\begin{acknowledgments}
We thank H. Boschker for the experimental data on critical temperatures and
for discussions. We are grateful to M. Cohen, A. Zunger and C. G. Van de Walle
for data on the deformation potential resulting from their works. This work
was supported by FWO-V projects G.0370.09N, G.0180.09N, G.0119.12N,
G.0122.12N, the WOG WO.035.04N (Belgium), the SNSF through Grant No.
200020-140761 and the National Center of Competence in Research (NCCR)
\textquotedblleft Materials with Novel Electronic Properties-MaNEP".
\end{acknowledgments}

\end{document}